\documentclass{article}

\usepackage{amssymb}
\usepackage{amsmath}
\usepackage{graphicx}
\usepackage[font=footnotesize,skip=20pt,width=1.2\linewidth]{caption}

\newcommand{\df}{\frac{}{}}
\newcommand{\tensor}{\otimes}
\newcommand{\wh}[1]{\widehat{#1}}

\renewcommand{\Re}{\text{Re}}
\newcommand{\NABLA}[1]{\nabla^{(#1)}}
\newcommand{\LAP}[1]{\text{Lap}^{(#1)}}
\newcommand{\DIV}[1]{\text{Div}^{(#1)}}
\newcommand{\EIN}[1]{\text{Ein}^{(#1)}}
\newcommand{\TR}[1]{\text{Tr}^{(#1)}}
\newcommand{\RIC}[1]{\text{Ric}^{(#1)}}
\newcommand{\MAN}[1]{\mathcal{M}^{(#1)}}
\newcommand{\TRREV}[1]{\mu^{(#1)}}

\newcommand{\CF}{{ \mathcal C}}
\newcommand{\PARTIAL[1]}{\partial_{{#1}} }

\newcommand{\RR}{\wh R(\tau)}
\newcommand{\ZZ}{\wh Z(\tau)}
\newcommand{\thth}{\wh \theta(\tau)}
\newcommand{\UU}{{ \mathcal U}}
\newcommand{\UUP}{{ \mathcal P}_{\mathcal U}}
\newcommand{\lrb}{\left(}
\newcommand{\rrb}{\right)}
\newcommand{\lsb}{\left[}
\newcommand{\rsb}{\right]}
\newcommand{\HMAX}{\mathcal{H}}

\newcommand{\CVRR}[1]{ {\mathcal R}^{ g_{#1}   }    }
\newcommand{\RRD}{\dot{\widehat {R}\,}(\tau)  } 
\newcommand{\ZZD}{\dot{\widehat {Z}\,}(\tau)  } 
\newcommand{\ththD}{\dot{\widehat {\theta}\,}(\tau)  } 
\newcommand{\RRDO}{\dot{\widehat{R}\,}(0)  } 
\newcommand{\ZZDO}{\dot{\widehat {Z}\,}(0)  } 
\newcommand{\ththDO}{\dot{\widehat {\theta}\,}(0)  } 
\newcommand{\RDO}{\dot{R}(0)  } 
\newcommand{\ZDO}{\dot{Z}(0)  } 
\newcommand{\thDO}{\dot{\theta}(0)  } 
\newcommand{\qquadand}{\qquad\text{and}\qquad}
\newcommand{\taumax}{\tau_{\text{max}}}
\newcommand{\AR}{\mathcal{A}(\taumax)}

\newcommand\WRINGA[2]{%
	\begin{figure}[!ht]
		\centering
		\includegraphics[width=#1\textwidth]{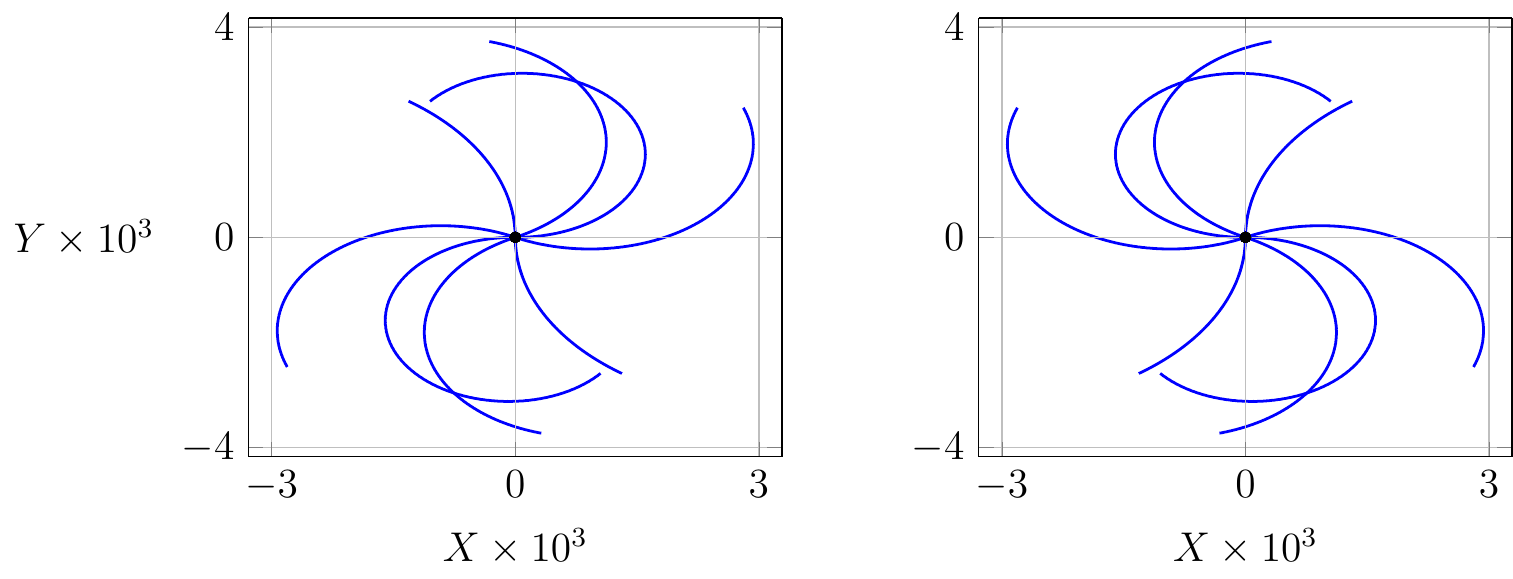}
		\caption{#2}
		\label{fig:wringa}
	\end{figure}
	}
\newcommand\WRINGB[2]{
	\begin{figure}[!ht]
		\centering
		\includegraphics[width=#1\textwidth]{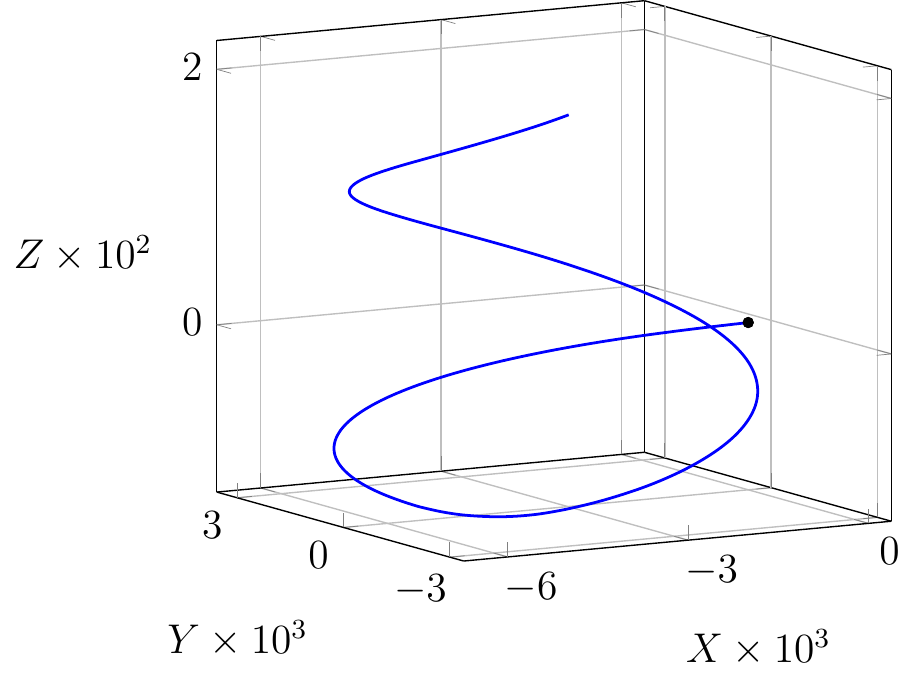}
		\caption{ #2 }
		\label{fig:wringb}
	\end{figure}
	}
\newcommand\WRINGC[2]{
	\begin{figure}[!ht]
		\centering
		\includegraphics[width=#1\textwidth]{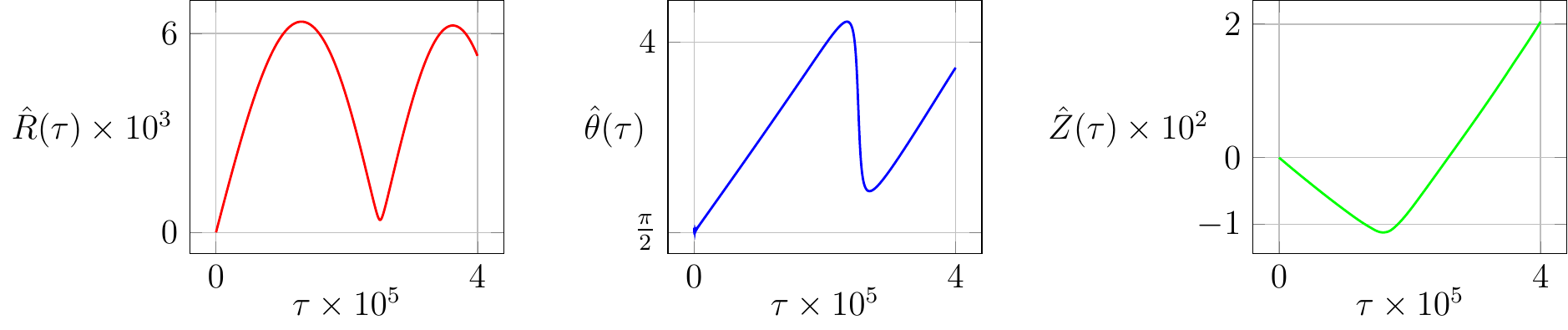}
		\caption{ #2 }
		\label{fig:wringc}
	\end{figure}
	}
\newcommand\WRINGD[2]{
	\begin{figure}[!ht]
		\centering
		\includegraphics[width=#1\textwidth]{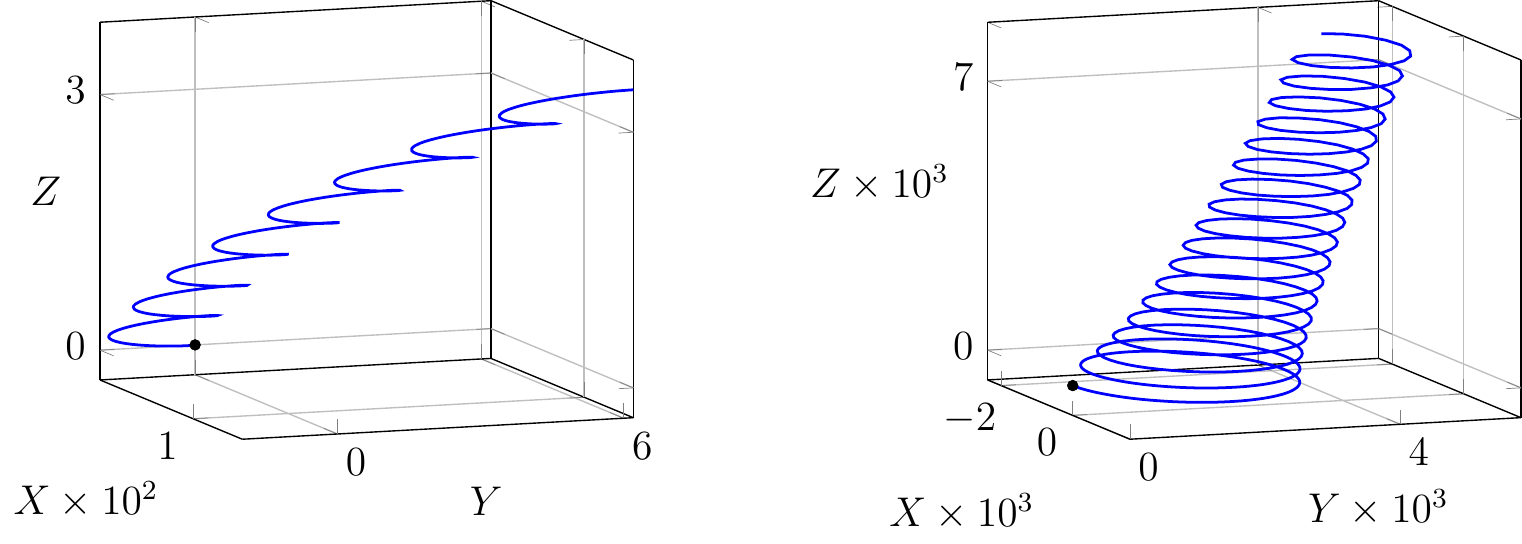}
		\caption{ #2 }
		\label{fig:wringd}
	\end{figure}
	}
\newcommand\WZLAYERS[2]{
	\begin{figure}[!ht]
		\centering
		\includegraphics[width=#1\textwidth]{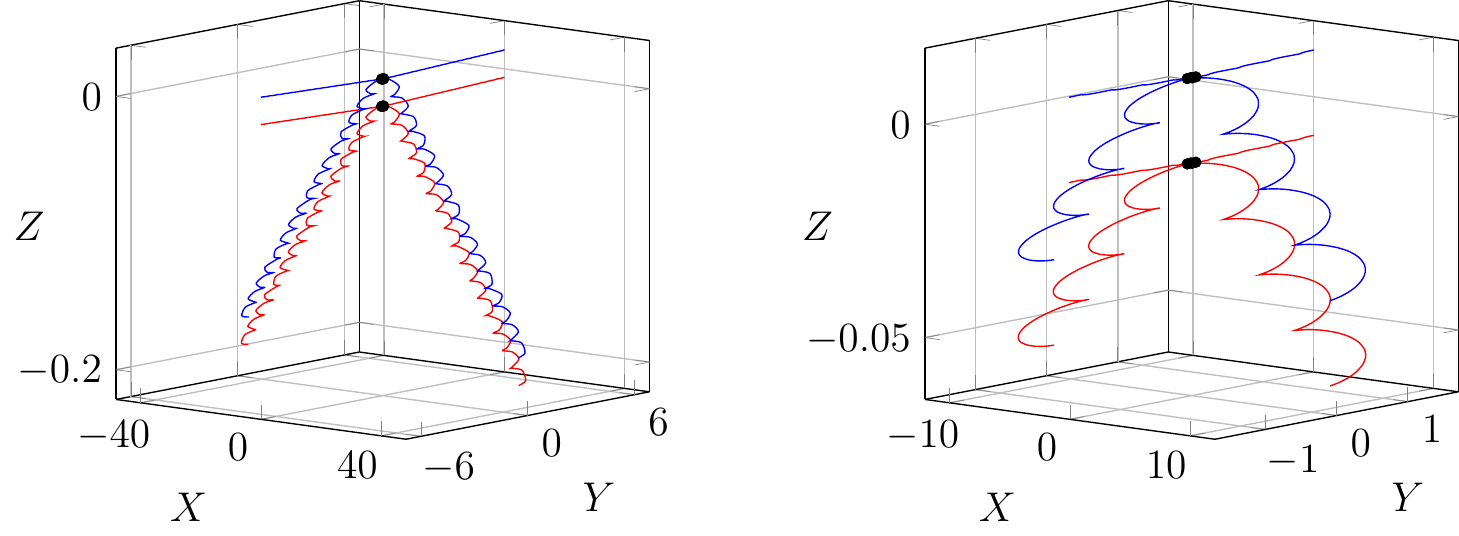}
		\caption{ #2 }
		\label{fig:wzlayers}
	\end{figure}
	}
\newcommand\WRINGE[2]{
	\begin{figure}[!ht]
		\centering
		\includegraphics[width=#1\textwidth]{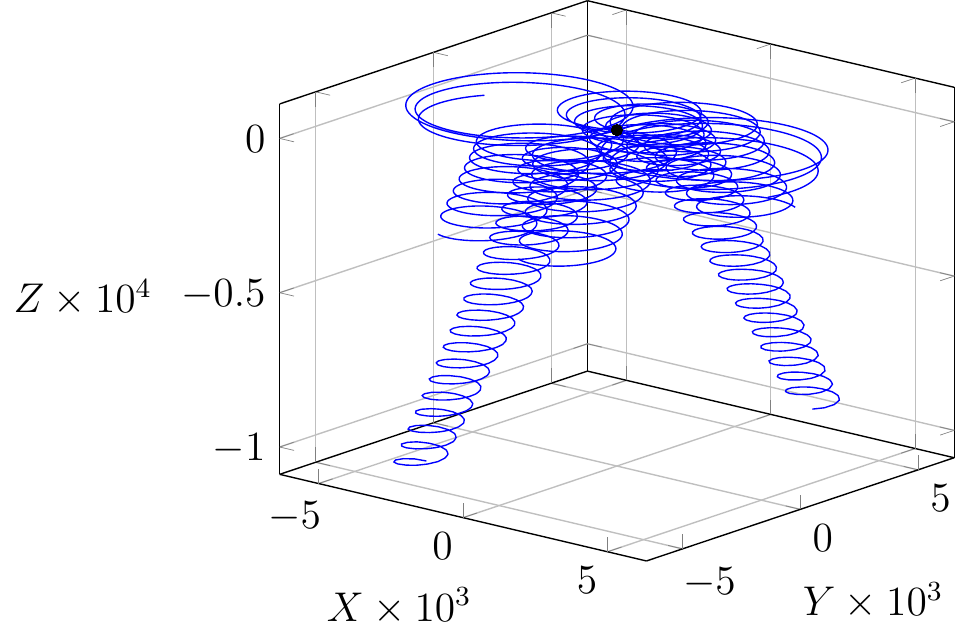}
		\caption{ #2 }
		\label{fig:wringe}
	\end{figure}
	}
\newcommand\HMAXRRZP[2]{
	\begin{figure}[!ht]
		\centering
		\includegraphics[width=#1\textwidth]{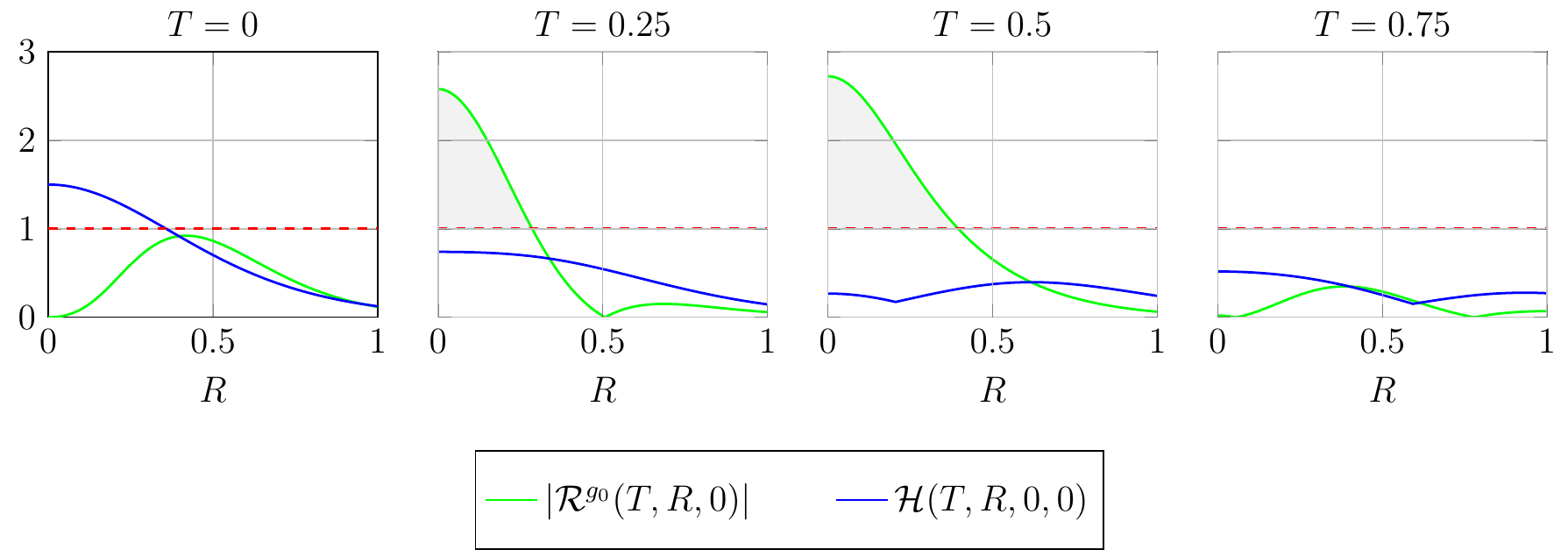}
		\caption{ #2 }
		\label{fig:hmaxrrzp}
	\end{figure}
	}
\newcommand\RRZP[2]{
	\begin{figure}[!ht]
		\centering
		\includegraphics[width=#1\textwidth]{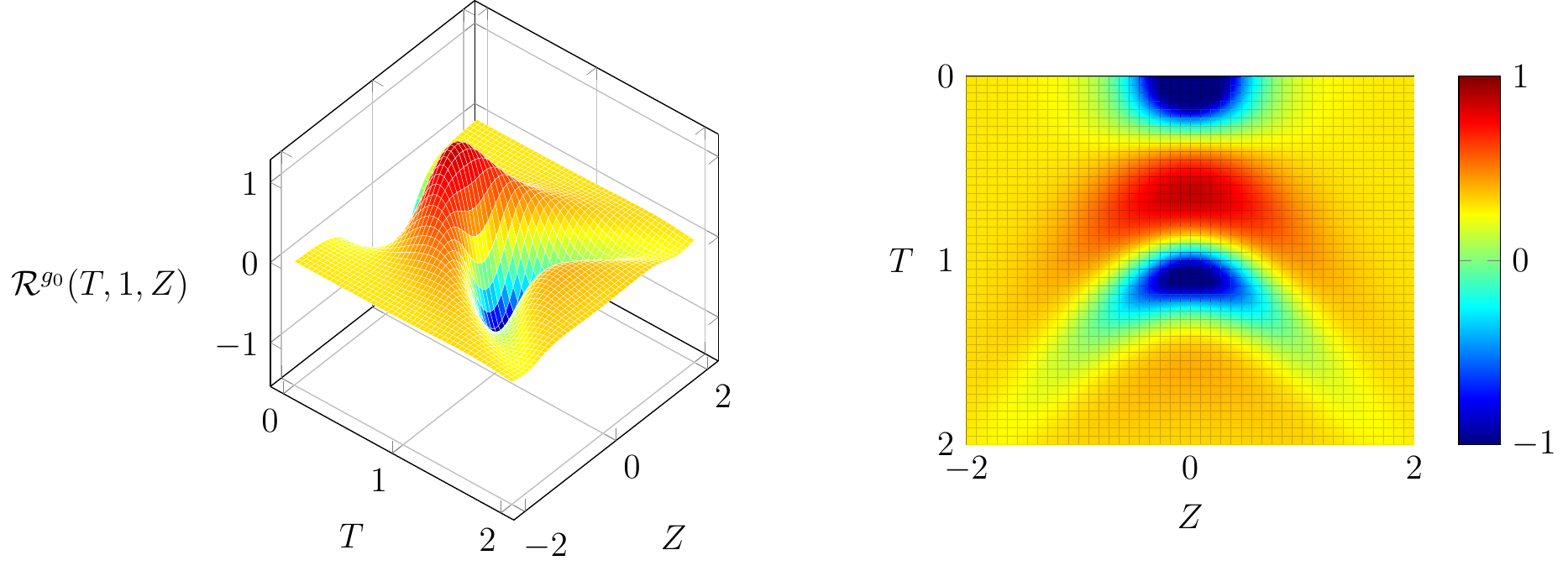}
		\caption{ #2 }
		\label{fig:rrzp}
	\end{figure}
	}
\newcommand\ZPZLAYERS[2]{
	\begin{figure}[!ht]
		\centering
		\includegraphics[width=#1\textwidth]{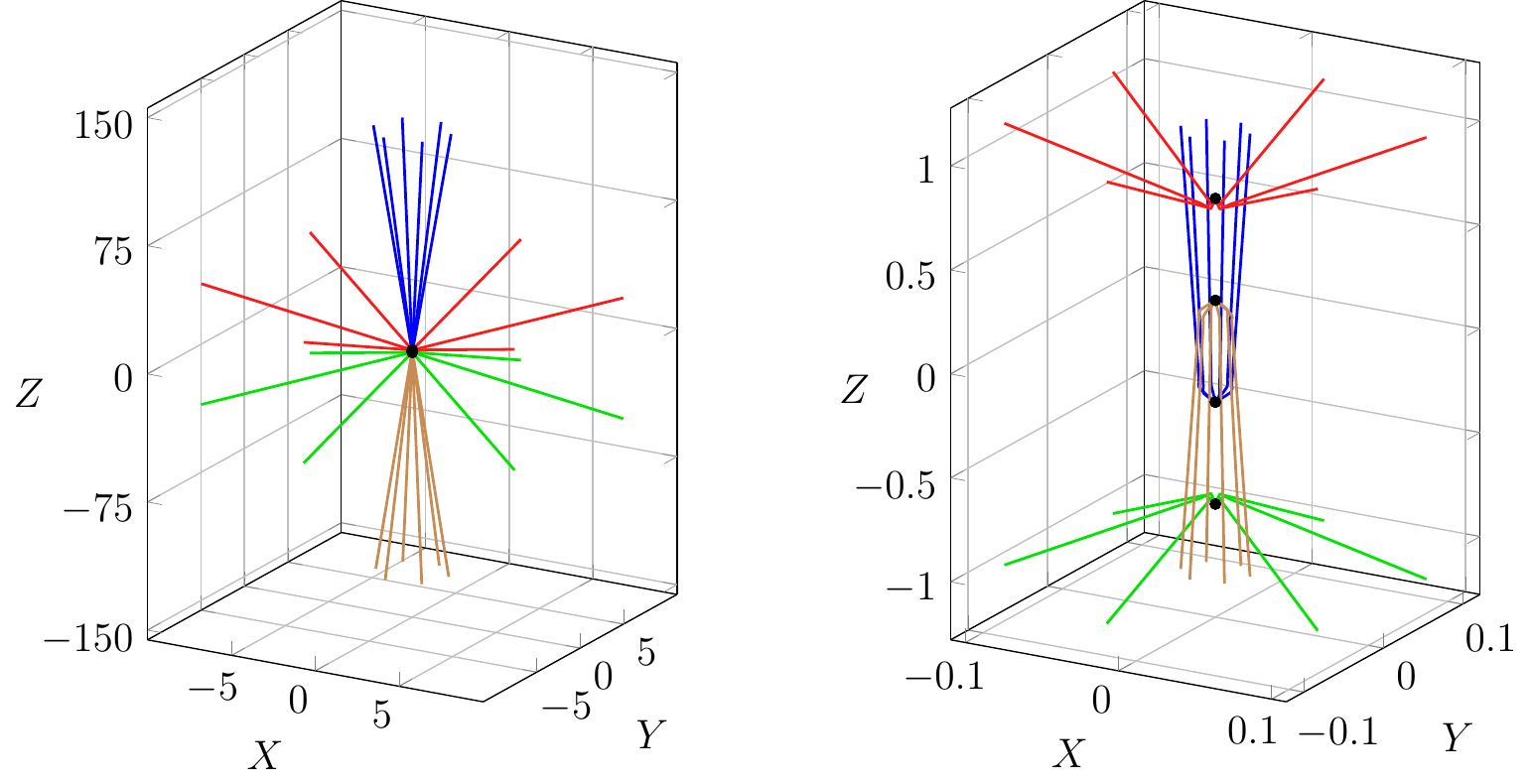}
		\caption{ #2 }
		\label{fig:zpzlayers}
	\end{figure}
	}
\newcommand\ZPZLAYERSQ[2]{
	\begin{figure}[!ht]
		\centering
		\includegraphics[width=#1\textwidth]{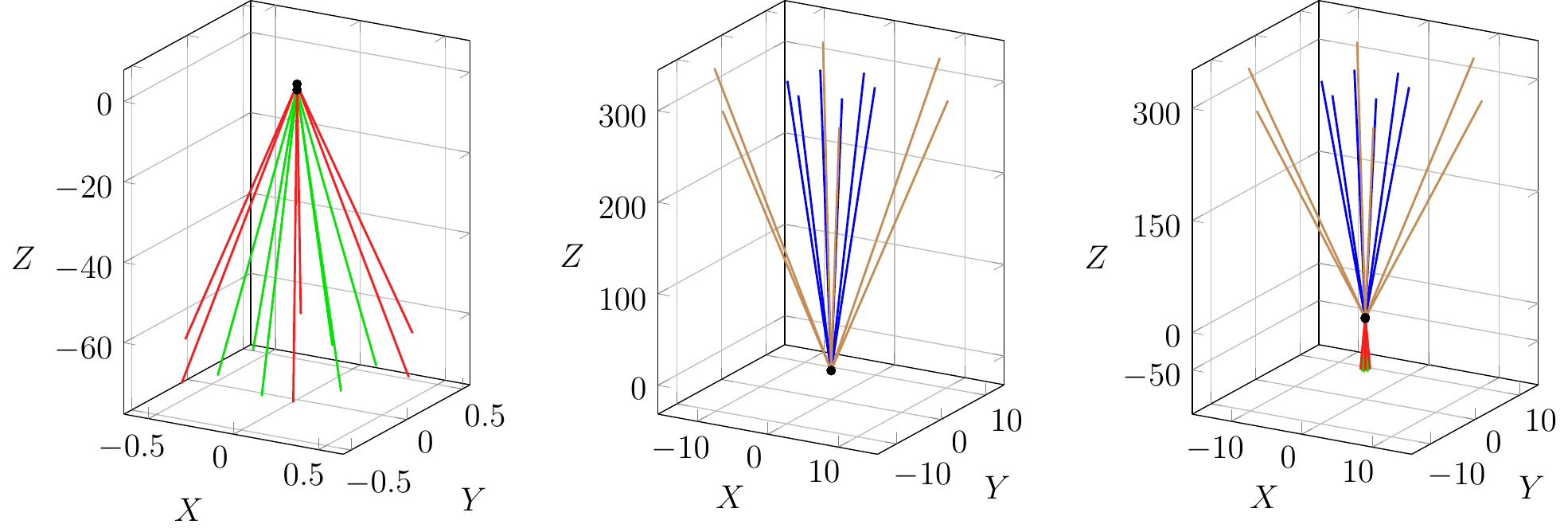}
		\caption{ #2 }
		\label{fig:zpzlayersq}
	\end{figure}
	}
	
\title{On Gravitational Chirality as the Genesis of Astrophysical Jets}
\author{R. W. Tucker${}^{1}$ and T. J. Walton${}^{2}$ \\[-0.2cm]
	\begin{tabular}{p{0.9\textwidth}} 
		\footnotesize ${}^{1}$ Department of Physics, University of Lancaster and Cockcroft Institute, Daresbury Laboratory, Warrington, UK \\[-0.6cm]
		\footnotesize${}^{2}$ Department of Mathematics, University of Bolton, Deane Campus, Bolton, UK
	\end{tabular}
 }
\begin{document}
\maketitle
\begin{abstract}
	It has been suggested that single and double jets observed emanating from certain astrophysical objects may have a purely gravitational origin. We discuss new classes of plane-fronted and pulsed gravitational wave solutions to the equation for perturbations of Ricci-flat spacetimes around Minkowski metrics, as models for the genesis of such phenomena. These solutions are classified in terms of their {\it chirality} and generate a family of non-stationary spacetime metrics. Particular members of these families are used as backgrounds in analysing time-like solutions to the geodesic equation for test particles. They are found numerically to exhibit {\it both} single and double jet-like features with dimensionless aspect ratios suggesting that it may be profitable to include such backgrounds in simulations of astrophysical jet dynamics from rotating accretion discs involving electromagnetic fields. 
\end{abstract}

\section{Introduction}
Many astrophysical phenomena find an adequate explanation in the context of Newtonian gravitation and Einstein's description of gravitation is routinely used (together with Maxwell's theory of electromagnetism and the use of  time-like spacetime geodesics to model the histories of  massive point test particles) to analyse a vast range of phenomena where non-Newtonian effects are manifest. However, there remain a number of intriguing astrophysical phenomena suggesting that our current understanding is incomplete. These include the large scale dynamics of the observed Universe and a detailed dynamics of certain compact stellar objects interacting with their environment. In this note we address the question of the dynamical origin of the extensive ``cosmic jets'' that have been observed emanating from a number of compact rotating sources. Such jets often contain radiating plasmas and are apparently the result of matter accreting on such sources in the presence of magnetic fields. One of the earliest models to explain these processes suggested that the gravitational fields of rotating black holes surrounded by a magnetised ``accretion disc'' could provide a viable mechanism \cite{blandford_z}. More recently, the significance of magneto-hydrodynamic processes in transferring angular momentum and energy into collimated jet structures has been recognised \cite{blandford,lynden,pringle}. Many of these models implicitly assume the existence of a magnetosphere in a {\it stationary} gravitational field and employ ``force-free electrodynamics'' in their development. To our knowledge, a dynamical model  that fully accounts for all the observed aspects of astrophysical jets does not exist. \\

However in recent years there has been mounting evidence, both theoretical and numerical, suggesting that the genesis of jets may have a purely gravitational origin. By the genesis of such phenomena we mean a mechanism that initiates the plasma collimation process whereby electrically charged matter  arises from initial distributions of neutral matter in a background gravitational field. In \cite{mashhoonPRD,mashhoonPLA}, the authors carefully analyse the properties of a class of Ricci-flat cylindrically symmetric spacetimes  possessing  time-like and null geodesics that  approach attractors confining massive particles to cylindrical spacetime structures.  Additional studies \cite{mashhoon_cosmic,mashhoon_peculiar,mashhoon_tidal} of the asymptotic behaviour of test particles on time-like geodesics with large Newtonian speeds relative to a class of co-moving observers  have given rise to the notion of {\it cosmic jets}  associated with different types of gravitational collapse scenarios  satisfying certain Einstein-Maxwell field systems. There has also been a recent approach based on certain approximations within a linearised gravitational framework involving ``gravito-magnetic fields'' generated by non-relativistic matter currents \cite{poirier}. All these investigations auger well for the construction of models for astrophysical jets that include  non-Newtonian gravitational fields  as well as electromagnetically induced plasma interactions. \\

Although astrophysical jets involve both gravitational and electromagnetic interactions with matter it is natural to explore the structure of electrically neutral test particle geodesics in non-stationary, anisotropic background metric spacetimes as a first approximation to what is undoubtedly a complex dynamical process. In this paper we explore geodesics that relax any assumption of a cylindrically Killing-symmetric background spacetime metric. Furthermore, to this end we construct particular exact solutions to the {\it linearised Einstein vacuum equations} which are then used to numerically calculate  time-like geodesics in non-stationary backgrounds. The use of the linearised Einstein vacuum equations facilitates the construction of  families of complex eigen-solutions with definite {\it chirality}  that are used to construct real spacetime metrics exhibiting families of time-like geodesics possessing particular jet-like characteristics on space-like hyper-surfaces.  Test particles on such time-like geodesics exhibit, in general, a well defined sense of ``handed-ness'' in space that we argue may offer a mechanism that initiates a flow of matter into directed jet-like structures. In particular, we construct families of plane-fronted gravitational wave metrics and  new non-stationary metrics having propagating {\it pulse-like} characteristics with bounded components in three-dimensional spatial domains. These are analogous to particular exact solutions of the vacuum Maxwell equations which we have recently shown can be used to model single-cycle electromagnetic laser pulses \cite{goto_lasers_JPA,goto_lasers_NIMB}. \\

In section~\ref{sect:LinEin}, our notation is established in the context of Einstein's vacuum equations on a spacetime manifold $\MAN{\wh{g}}$ with a Lorentzian metric $\wh{g}$ and their linearisation about a flat Minkowski metric $\eta$ on a region $\mathcal{U}\subset\MAN{\wh{g}}$. \\

Section~\ref{sect:GravWaves} describes a family of complex plane-fronted gravitational wave exact solutions to the linearised field equations that are eigen-tensors of the  complex axial-symmetry operator about any particular direction in space. We call the eigenvalues of this operator ``chirality'' and show how the associated ``helicity'' of the real-part solutions with respect to their direction of propagation is correlated with certain properties of time-like geodesics associated with the real linearised spacetime metric.  The simplest {\it harmonic} plane-fronted  gravitational wave modes have helicity $\pm 2$  and, in general, induce a (modulated) non-circular helicoidal motion of massive test-particles that are initially arranged (with non-zero speeds) in a ring that lies in a plane orthogonal to the direction of propagation of such  waves. The set  composed of the individual particle paths form a {\it uni-directional} bi-jet-like array in space emanating from such a ring in a background harmonic wave of definite helicity.\\

By contrast in section~\ref{sect:compact}, we give a construction of exact solutions to the linearised equations with pulse-like perturbations that are  bounded in all spatial dimensions. The simplest gravitational pulse-like solutions have chirality zero and are shown  to generate a real non-singular Ricci curvature scalar field on $\mathcal{U}$ with well defined loci in spacetime emanating from  a core  containing a maximum at a particular event. Future-pointing time-like geodesics that emanate from events on rings containing massive test particles centred on this core can give rise to single and {\it oppositely directed}  jet-like arrays in space, transverse to the plane of the rings (bipolar outflows).\\

It is assumed that exact analytic solutions to the linearised Einstein vacuum equations and their associated  time-like geodesics  will exhibit features that persist to some degree beyond the linearisation regime and, in particular, offer an approach to a better understanding of the genesis of observable cosmic jets  in models that include charged matter with plasma interactions. Based on a numerical exploration of particular time-like geodesics associated with background metrics constructed from complex eigen-solutions with definite chirality, we conclude  that it may be profitable to include these non-Newtonian  gravitational backgrounds  in simulations of  cosmic jet dynamics  from rotating accretion discs involving electromagnetic fields. \\

\newpage
\section{The Linearised Einstein Vacuum System} 
\label{sect:LinEin}
In any matter free domain of spacetime $\mathcal{U}\subset\MAN{\wh{g}}$, an Einsteinian gravitational field is described by a symmetric covariant rank-two tensor $\wh{g}$ with Lorentzian signature that satisfies the vacuum Einstein equation $\EIN{\wh{g}}=0$ where
\begin{eqnarray}\label{VAC}
	\EIN{\wh{g}} &\;\equiv\;& \RIC{\wh{g}} - \frac{1}{2}\,\TR{\wh{g}}(\RIC{\wh{g}})\,\wh{g}
\end{eqnarray}
and $\RIC{\wh{g}}$ is the Levi-Civita Ricci tensor associated with the torsion-free, metric-compatible Levi-Civita connection $\NABLA{\wh{g}}$. A coordinate independent linearisation of (\ref{VAC}) can be found in \cite{stewart,tuckerGEM}. In particular a linearisation about a flat Minkowski spacetime metric $\eta$ on $\mathcal{U}$ determines the linearised metric\footnote{%
	Throughout this paper we assign physical dimensions of length${}^{2}$ to the {\it tensors} $g$ and $\eta$. The Ricci curvature scalar associated with $g$ then has the dimensions of length${}^{-2}$. In a $g$-orthonormal coframe the components of $g$ are $\{-1,1,1,1\}$ and in an $\eta$-orthonormal coframe the components of $\eta$ are $\{-1,1,1,1\}$. This does not imply that components of  the tensor field $\eta$ are necessarily constant in an arbitrary coframe on $\UU$.%
	} 
$g=\eta + h$ and to first order one writes $\wh{g} = g + O(\kappa^{2}).$  The variable  $\kappa$ is a parameter in $h$ used to keep track of the expansion order and
\begin{eqnarray}\label{ETA}
	\eta &\;=\;& -\, e^{0} \tensor e^{0} + \sum_{k=1}^{3} e^{k} \tensor e^{k} \;=\; \eta_{ab}\,e^{a} \tensor e^{b} \qquad  \text{where} \quad a,b=0,1,2,3
\end{eqnarray}
in {\it any}  $\eta$-orthonormal coframe\footnote{%
	An arbitrary local coframe  is a set of $1$-forms  $\{e^{a}\}$ on $\UU$  satisfying $e^{0}\wedge e^{1}\wedge e^{2}\wedge e^{3} \neq 0 $.  If $\psi=\psi_{ab}e^{a}\tensor e^{b}$ in {\it any}  such coframe, $\TR{\eta}\psi=\psi_{ab}\eta^{ab}$ with $\eta^{ab}\eta_{bc}=\delta^{a}_{c}$ in terms of the Kronecker symbols $\delta^{a}_{c}$.%
	} 
on $\UU$. Since we explore the source-free Einstein equation (relevant to the motion of test-matter far from any sources) the physical scale associated with any linearised solutions must be fixed by the solutions themselves rather than any coupling to self-gravitating matter. Furthermore since only dimensionless relative scales have any significance we define the tensor $h$ to be a {\it perturbation} of $\eta$ on $\UU$ relative to  {\it any} local $\eta$-orthonormal coframe $\{e^{a}\}$  provided 
\begin{eqnarray}\label{CNDS}
	|\,h(X_{a},X_{b})\,| \;<\; 1 \qquad \text{on $\mathcal{U}$}  \qquad  \text{for all $a,b=0,1,2,3$}, 
\end{eqnarray} 
where  $e^{a} \in T^{*}\mathcal{U}$, $X_{b} \in T\mathcal{U}$, $e^{a}(X_{b})=\delta^{a}_{b}$. It should be noted that the perturbation order of the $\eta$-covariant {\it derivative} of a tensor and its $\eta$-trace relative to such a coframe is not necessarily of the same order as that assigned to the tensor. Thus perturbation order is not synonymous with ``scale'' in this context. We use the conditions (\ref{CNDS}) to {\it define} perturbative Lorentzian spacetime  domains $\UUP\subset \UU$ to be regions where 
\begin{eqnarray*}
	\mathop{\text{max}}_{0\leq a,b\leq 3} \,\vert h(X_{a},X_{b})\vert &\;<\;& 1.
\end{eqnarray*}
The real tensor $h\equiv\TRREV{\eta}{\psi'}$ with $\psi'\equiv\Re(\psi)$  may be constructed from any complex covariant symmetric rank-two tensor $\psi$ satisfying \cite{tuckerGEM}:
\begin{eqnarray*}
	\LAP{\eta}\psi - 2\TRREV{\eta}\lrb\text{Sym}\NABLA{\eta}(\DIV{\eta}\psi) \rrb &\;=\;& 0.
\end{eqnarray*}
Here and below, $\NABLA{\eta}$ denotes the operator of Levi-Civita covariant differentiation associated with $\eta$,  $X^{a}\equiv\eta^{ab}X_{b}$, $Y\equiv\NABLA{\eta}_{X_{a}}X^{a}$ and for all covariant symmetric rank-two tensors  $T$ on $\UU$: 
\begin{eqnarray*}
	\LAP{\eta}T		&\;\equiv\;& \NABLA{\eta}_{Y}T - \NABLA{\eta}_{X_{a}}\NABLA{\eta}_{X^{a}}T	\\[0.1cm]
	\DIV{\eta}T		&\;\equiv\;& (\NABLA{\eta}_{X_{a}}T)(X^{a},-) \\[0.1cm]
	\TRREV{\eta}T	&\;\equiv\;& T - \frac{1}{2}(\TR{\eta}T)\,\eta . 
\end{eqnarray*}
Since for any $g$ the reverse-trace map $\TRREV{g}$ satisfies $\TRREV{g}\TRREV{g}=\text{Id}$, if $\psi'$ is trace-free with respect to $\eta$, then $h=\psi'$. If $\psi$ is also divergence-free with respect to $\eta$, then $\LAP{\eta}\psi=0$. Thus, divergence-free, trace-free solutions $\psi$ satisfy:
\begin{eqnarray}\label{DFTF_LAP_EQ}
	\LAP{\eta}\psi \;=\; 0 \qquadand	\DIV{\eta}\psi \;=\; 0.
\end{eqnarray}
Given $\psi$ and hence $g=\eta + h$, all proper-time parametrised time-like spacetime geodesics $C$ on $\mathcal{U}$, with tangent vector $\dot{C}$, associated with $g$,  must satisfy the differential-algebraic system
\begin{eqnarray}
	\label{NORM}	g(\dot{C},\dot{C}) &\;=\;& -1 \\[0.2cm]
	\label{GEOD_EQ}	\NABLA{g}_{\dot{C}}\dot{C} &\;=\;& 0.
\end{eqnarray}	 
If  any  worldline $C$   has components $C^{\mu}(\tau)$ in any local chart on $\mathcal{U}$ with coordinates $\{x^{\mu}\}$  and  $\dot{C}^{\mu}\equiv\partial_{\tau} C^{\mu}$ then
\begin{eqnarray}\label{ACCEL_EQ}
	\NABLA{g}_{\dot{C}}\dot{C} \;\equiv\; \frac{D \dot{C}^{\mu}}{d\tau} \partial_{\mu} &\;=\;& 
   \lrb\frac{ d\dot{C}^{\mu}}{d\tau}+ ( \Gamma^{\mu}_{\alpha\beta} \circ C) \, \dot{C}^{\alpha}\dot{C}^{\beta}\rrb\, \partial_{\mu}
\end{eqnarray}
where $\Gamma^{\mu}_{\alpha\beta}$ denotes a Christoffel symbol associated with $\NABLA{g}$.\\

In the following sections only solutions to (\ref{NORM}) and (\ref{GEOD_EQ}) that lie in the perturbative domains $\UUP$ are displayed. The worldline of an idealised {\it observer} in $\mathcal{U}$ is modelled by the integral curve $C_{V}$ of a future-pointing time-like unit vector field $V$, (i.e. $g(V,V)=-1$).  At any event in  $\mathcal{U}$ the $g$-orthogonal decomposition of $\dot{C}$ with respect to an observer $C_V$:
\begin{eqnarray*}
	\dot{C} &\;=\;& \boldsymbol{\nu} - g(\dot{C},V) V,       
\end{eqnarray*}
with $g(\boldsymbol{\nu},V)=0$ defines the Newtonian 3-velocity field $\mathbf{v}$ on $C$ relative to the integral curve $C_{V}$  that it intersects in spacetime: 
\begin{eqnarray*}
	\mathbf{v} &\;=\;& \frac{\boldsymbol{\nu}}{g(\dot{C},V)} \;\equiv\; \frac{\dot{C}}{g( \dot{C},V)} + V. 
\end{eqnarray*}
Relative to $C_{V}$, the observed ``Newtonian speed''  of the proper-time parameterised worldline $C$  at any event $C^{\mu}(\tau)$ is then $v\equiv\sqrt{g(\mathbf{v} , \mathbf{v} )\,  }$. If $\NABLA{g}_{V}V=0$ the observer is said to be geodesic otherwise it will be accelerating. If there exists a local coordinate system $\{t,\xi_{1},\xi_{2},\xi_{3}\}$ on $\mathcal{U}$  with $\partial_{t}$  time-like, in which $C_{V}$ can be parameterised  monotonically with $\lambda$ as $t=\lambda$, $\xi_{1}=\xi_{1}(0)$, $\xi_{2}=\xi_{2}(0)$, $\xi_{3}=\xi_{3}(0)$ such an observer is said to be at rest in this coordinate system. Although {\it any} particular time-like worldline defines a local ``rest observer'' in {\it some chart}, only the existence of a {\it family} of rest observers in a particular chart $\Phi_{\mathcal{U}}$ on $\mathcal{U}$ provides a way to interpret the Newtonian velocity of any event on a time-like worldline that is not necessarily a rest-observer in $\mathcal{U}$. In units\footnote{%
	We introduce a length scale parameter $L_{0}$ to relate coordinates $\{t,r,z\}$  
	and angular frequency $\omega$ with physical dimensions $\{\text{time,length,length}\}$  and 1/time respectively to the dimensionless variables $\{T,R,Z\}$ and $\Omega$ so that $r = L_{0}R$, $z = L_{0}Z$, $c_{0}t = L_{0}T$ and $\omega = \Omega c_{0}/L_{0}$, where $c_{0}$ is a fundamental constant with the physical dimensions of speed. In SI units $c_{0}=3\times 10^{8}$ m/sec. The dimensionless amplitudes in (\ref{Lq}) define amplitudes $L_{0}q_{1}$ and $L_{0}q_{2}$ with physical dimensions of length. In this scheme with $g(\dot{C},\dot{C})=-1$, the parameter $\tau$ has length dimensions and its conversion to a parameter $\tau^{\prime}$ with dimensions of a clock time is given by $\tau^{\prime}=\tau/c_{0}=\wh{\tau} L_{0}/c_{0}$  where $\wh{\tau}$ is a dimensionless parameter.%
} 
with $c_{0}=1$, a point particle of rest-mass $m_{0}$ with a worldline $C$, when observed by $C_{V}$,  has energy and 3-momentum  values at any event on $C$ given by    $\mathcal{E}_{V} = \gamma_{V} m_{0}$ and $\mathbf{p}_{V}=\gamma_{V} m_{0}\mathbf{v}$ respectively, where $\gamma_{V}\equiv 1/\sqrt{ 1- g( \mathbf{v}, \mathbf{v} )\,}$.\\

\newpage
\section{Gravitational Waves}
\label{sect:GravWaves}
The recent experimental detection of gravitational waves lends credence to the existence of non-stationary metrics on spacetime domains in the vicinity of colliding black holes. There are good reasons to believe that such domains may exist in the vicinity of other non-stationary astrophysical processes. In this section we explore the properties of particular  proper time parameterised plane-fronted gravitational wave spacetimes in the perturbative domains $\UUP$ defined above. \\

In a local chart $\Phi_{\mathcal{U}}$ possessing dimensionless coordinates $\{T,R,\theta,Z\}$ with $T\geq 0$, $R>0$, $\theta\in[0,2\pi)$ and $|Z|\geq 0 $ on a spacetime domain $\mathcal{U}\subset \mathcal{M}^{(\wh{g})}$, a local coframe $\CF$ adapted to these co-ordinates is $\{e^{0}=dT,\, e^{1}=dR, \,e^{2}=R\,d\theta, \,e^{3}=dZ\}$ in terms of the exterior derivative operator $d$.  With $\eta$ given by (\ref{ETA}), this coframe is $\eta$-orthonormal but not in general $g$-orthonormal. Such a chart facilitates the coordination of a series of massive test particles initially arranged  in a series of concentric rings with different values of $R$ lying in spatial planes with different values of $Z$ at $T=0$. Furthermore, we define on $\UU$:
\begin{eqnarray*}
	\HMAX(T,R,\theta,Z) &\;\equiv\;& \mathop{\text{max}}_{0\leq a,b\leq 3} \,\vert h(X_{a},X_{b})\vert.
\end{eqnarray*}
For $\sigma= \pm 1$, $m_{1}, m_{2}\in\mathbb{Z}$, particular non-harmonic complex plane-fronted gravitational wave solutions to (\ref{DFTF_LAP_EQ}) take the form
\begin{eqnarray}\label{Psi_GW}
	\psi^{\sigma}_{m_{1} + m_{2}}(T,R,\theta,Z) &\;=\;& W(Z-\sigma T)\; \text{Sym}\lrb d\,( R^{|m_{1}|} \,e^{im_{1}\theta} \,) \tensor d\, (R^{|m_{2}|} \,e^{im_{2}\theta} \,) \rrb
\end{eqnarray}
where $W$ is an arbitrary complex-valued function that is at least twice differentiable on $\mathcal{U}$. Waves with opposite values of $\sigma$ propagate in opposite directions along the axis $Z$. These tensors satisfy\footnote{%
	For any vector field $\xi$, the operator $\mathcal{L}_{\xi}$ denotes Lie differentiation with respect to $\xi$.%
	}%
\begin{eqnarray*}
	\frac{1}{i} \mathcal{L}_{\partial_{\theta}}\psi^{\sigma}_{m_{1} + m_{2}} &\;=\;& (m_{1} + m_{2})\,\psi^{\sigma}_{m_{1}+m_{2}}
\end{eqnarray*}
and will be called complex eigen-tensors of definite chirality $m_{1} + m_{2}$. Such tensors give rise to real tensors $(\psi^{\sigma}_{m_{1} + m_{2}})' \equiv \text{Re}(\psi^{\sigma}_{m_{1} + m_{2}})$ with ``helicity'' $(m_{1} + m_{2})\sigma$. If $ |m_{1}| \geq 1$,  $|m_{2}|\geq 1$ and $m_{1} m_{2} >0$,  it may be shown that $ \LAP{\eta} \psi^{\sigma}_{m_{1} + m_{2}}=0, \,\, \DIV{\eta} \psi^{\sigma}_{m_{1} + m_{2}}=0$ and $ \TR{\eta} \psi^{\sigma}_{m_{1} + m_{2}}=0  $. Since $\LAP{\eta}$ and $\DIV{\eta}$ are linear operators, the superposition $\sum_{\sigma,n}C^{\sigma}_{n}\psi^{\sigma}_{n}$ with arbitrary complex constant coefficients $C^{\sigma}_{n}$ is also a solution of (\ref{DFTF_LAP_EQ}). From the  simplest  trace-free solution with $m_{1}=1$ and $m_{2}=1$  one may construct from  $(\psi^{\sigma}_{2})'$  the  real  helicity $2\sigma$ symmetric covariant metric tensor $g^{\sigma}_{2}$ on $\mathcal{U}$:
\begin{eqnarray*}
	g^{\sigma}_{2} &\;\equiv\;& \eta + (\psi^{\sigma}_{2})'.
\end{eqnarray*}
Pure {\it harmonic} plane-fronted gravitational waves arise when the function $W$ in (\ref{Psi_GW}) is a bounded trigonometric function of its arguments.  The higher complex chirality solutions are related to  $\psi^{\sigma}_{2}$ by the formula:
\begin{eqnarray*}
	\psi^{\sigma}_{m_{1} + m_{2}} &\;=\;&  2m_{1}m_{2} \,e^{i(m_{1} + m_{2}-2)\theta} \, R^{ m_{1} + m_{2}-2 }\psi^{\sigma}_{2}
\end{eqnarray*}
The real tensors  $(\psi^{\sigma}_{2})^{\prime}$ remain trace-free, divergence-free solutions to $\LAP{\eta}(\psi^{\sigma}_{2})^{\prime}=0$ under the finite real rotation map  $\theta\mapsto \theta +\theta_{0}$ with arbitrary real constant  $\theta_{0}$.\\

We have  explored  numerically the  structure of time-like geodesics on $\mathcal{U}$ associated with $g^{\sigma}_{2}$ for pure harmonic plane-fronted gravitational waves where: 
\begin{eqnarray}\label{Lq}
	W(u) &\;\equiv\;& q_{1}\,\cos(\Omega u) + i \,q_{2}\,\sin(\Omega u)
\end{eqnarray}
with dimensionless angular frequency $\Omega$ and real dimensionless  amplitudes $q_{1},q_{2}$. In these gravitational wave spacetimes the geodesic vector field $\partial_{T} \in T\mathcal{U}$ satisfies $ g^{\sigma}_{2}(\partial_{T},\partial_{T})=-1$. In $\mathcal{U}$, the  metric tensor $g^{1}_{2}$ has non-zero  components in the coframe $\CF$:
\begin{eqnarray*}
	g_{00} &\;=\;& -1 \\[0.1cm]
	g_{11} &\;=\;& 1 + \frac{1}{2}\lsb\df(q_{1} +q_{2}) \cos\lrb\df \Omega(T-Z) - 2\theta \rrb   + (q_{1} - q_{2}) \cos\lrb \df  \Omega(T-Z) + 2\theta \rrb \rsb \\[0.1cm]
	g_{12} &\;=\;& g_{21} \;=\; \frac{1}{2}\lsb\df(q_{1} +q_{2}) \sin\lrb\df \Omega(T-Z) - 2\theta  \rrb   + (q_{2} - q_{1}) \sin\lrb\df \Omega(T-Z) + 2\theta  \rrb  \rsb \\[0.1cm]
	g_{22} &\;=\;& 1 + \frac{1}{2}\lsb\df-(q_{1} +q_{2}) \cos\lrb\df \Omega(T-Z) - 2\theta  \rrb   + (q_{2} - q_{1}) \cos\lrb\df \Omega(T-Z) + 2\theta  \rrb \rsb\\[0.1cm]
	g_{33} &\;=\;& 1.
\end{eqnarray*}
The integral curves of the vector field $\partial_{T}$ provide, in $\Phi_{\mathcal{U}}$, a family of  geodesic rest observers on $\mathcal{U}$. In this chart we write the proper-time parameterised  geodesic curves  as $\{L_{0} T=\tau,\, R=\wh{R}(\tau), \, \theta=\wh{\theta}(\tau), \, Z=\wh{Z}(\tau)\}$. Only three of the equations in (\ref{ACCEL_EQ})  are independent. In the $L_{0}T=\tau$ gauge the equations with $\mu=1,2,3$ are solved numerically for $\wh{R}(\tau)$, $\wh{\theta}(\tau)$, $\wh{Z}(\tau)$ with prescribed values of $\wh{R}(0)$, $\wh{\theta}(0)$, $\wh{Z}(0)$  and $\RRDO$, $\ththDO$, $\ZZDO$. These initial conditions must be chosen consistently with the condition $g^{\sigma}_{2}(\dot{C},\dot{C})\vert_{\tau=0}=-1$  which is then preserved for all $\tau$.  Furthermore  the parameters $q_{1},q_{2},\Omega$ are chosen so that throughout the numerical integration  of (\ref{NORM}) and (\ref{GEOD_EQ}) {\it all components} of $(\psi^{\sigma}_{2})^{\prime}$ in the $\eta$-orthogonal coframe evaluated on $C$ for $\tau\geq 0$ remain less than unity in absolute value. This ensures that all computed geodesics  lie in  the  perturbative domains  $\UUP$.  The solutions to (\ref{NORM}) and (\ref{GEOD_EQ}) when $\RRDO=0$, $\ththDO=0$, $\ZZDO=0$ are  $\wh{R}(\tau)=\wh{R}(0)$, $\wh{\theta}(\tau)=\wh{\theta}(0)$, $\wh{Z}(\tau)=\wh{Z}(0)$, i.e. test particles on these geodesics remain at rest relative to rest observers. Solutions $R(\tau)$, $\theta(\tau)$, $Z(\tau)$ are displayed  individually or as  single space-curves in Cartesian domains  labelled with dimensionless coordinates $X,Y,Z$ where $X=R \cos(\theta)$ and $Y=R\sin(\theta)$. For each solution in $\UUP$ one can calculate the square of the Newtonian speed $v$ at any $\tau$ relative  to a rest observer:
\begin{eqnarray*}
	v(\tau)^{2}	&\;=\;& \RRD^{2}  + \ZZD^{2} + \RR^{2}\,\ththD^{2} \\[0.1cm]
				& & + \; \RR\,\RRD\,\ththD\, \lsb \df (q_{1}+q_{2})\sin\!\lrb\df\!\phi_{-}(\tau)\rrb + (q_{2}-q_{1})\sin\!\lrb\df\!\phi_{+}(\tau)\rrb\rsb\\[0.1cm]
				& & + \; \frac{1}{2}\lrb\RRD^{2} - \RR^{2}\,\ththD^{2}\rrb\! \lsb (q_{1}+q_{2})\cos\!\lrb\df\!\phi_{-}(\tau)\rrb + (q_{1}-q_{2})\cos\!\lrb\df\!\phi_{+}(\tau)\rrb\rsb
\end{eqnarray*}
where
\begin{eqnarray*}
	\phi_{\pm}(\tau) &\;\equiv\;& \Omega\lrb\df\tau-\ZZ\rrb \pm 2\thth.
\end{eqnarray*}
\WRINGA{0.8}{%
	Eight time-like geodesics all evolving from  proper time $\tau=0$ to $\tau=4 \times 10^{4}$ from eight equally  arranged initial  locations around a ring of radius 1 unresolved in these figures) with additional initial conditions  $Z(0) = 0$, $\RDO = 0.1$, $\ZDO = 0$, $\thDO = 0$. The parameters of the helicity $\pm2$  background gravitational wave metric are $L_{0}=1$, $q_{1}=q_{2} =1/3$, $\Omega=1/20$.  The sense of the helicoidal motion for helicity 2 shown on the left is reversed for waves of opposite helicity -2 shown on the right.
	}
To display the general features of the geodesic solutions to (\ref{DFTF_LAP_EQ}) that exhibit jet-like features for specified initial conditions we select particular values of the parameters $\Omega, q_{1},q_{2}, L_{0}$ and helicity $2\sigma$ where $\sigma= \pm 1$. Having the notion of a thick accretion disc in mind, we choose initial ($\tau=0$)  positions of individual geodesics to be equally spaced in $\theta$ on rings with fixed $R$ in planes with fixed $Z$. On the left in figure~\ref{fig:wringa} a set of 8 such time-like geodesics is  displayed in an $X$-$Y$ projection, each  for an evolution proper time from $\tau=0$ to $\tau=4\times 10^{4}$. Each geodesic exhibits a segment of  a helicoidal structure in a background gravitational wave of definite helicity 2. The sense of the helicoidal motion is reversed for waves of opposite helicity as demonstrated in the figure on the right. Although the average motion of the helicoids along the $Z$-axis is determined by the direction of the wave, the motions $\RR$ and $\ZZ$ of a test particle with  an initial non-zero radial speed are not immediately monotonic in $\tau$, as evident in figure~\ref{fig:wringb}.%

\WRINGB{0.5}{%
	A single  time-like geodesics evolving from  $\tau=0$ to $\tau=4 \times 10^{5}$ from an initial  location (denoted by a black dot) at $\theta=\pi/2$  on a ring of radius 1 with the same initial conditions as in figure~\ref{fig:wringa}:   $ Z(0) = 0$, $\RDO = 0.1$, $\ZDO = 0$, $\thDO = 0$ and same parameters of the helicity 2  background gravitational wave metric with  $L_{0}=1$, $q_{1}=q_{2} =1/3$, $\Omega=1/20$. Components of this space-curve, displayed in figure~\ref{fig:wringc}, indicate that the axial motion is not monotonic in proper-time.\newline%
	}
The corresponding parametric behaviours of the solutions $\RR,\thth,\ZZ$ are displayed in figure~\ref{fig:wringc}. \\
  
\WRINGC{0.9}{%
	Parametric representations of the space-curve shown in figure~\ref{fig:wringb}. Note that the axial motion is not monotonic in proper-time.  %
	}     
	
The left hand side plot in figure~\ref{fig:wringd}  displays a short segment of a typical helicoid  on a scale that makes visible a cycloidal-like oscillatory modulation. The spatial periodicity of this modulation is determined by the value of $\Omega$ and decreases as this parameter increases. The right hand side of figure~\ref{fig:wringd} displays on a lower resolution (thereby suppressing the cycloidal modulation) a segment of a typical helicoid that exhibits helicoidal drift and a decrease of helicoidal radius as $\tau$ increases.%
\WRINGD{0.8}{%
	The left hand side plot displays a short segment of a typical  geodesic helicoid  on a scale where $\tau$ varies from zero to $\tau=1 \times 10^{3}$ and thereby makes visible  a cycloidal-like oscillatory modulation. The spatial periodicity of this modulation is determined by the value of $\Omega$ and decreases as this parameter increases. The right hand side plot   displays  on a lower resolution, where $\tau$ varies from zero to $\tau=5 \times 10^{5}$ (thereby suppressing the visibility of the cycloidal modulation),  a segment of a typical geodesic helicoid that exhibits helicoidal drift and a decrease of helicoidal radius as $\tau$ increases. The parameters of the helicity 2  background gravitational wave metric are $ L_{0}=1$, $q_{1}=q_{2} =1/3$, $\Omega=1/20$  and initial conditions (denoted by black dots) $R(0)=1$,  $Z(0) = 0$,  $\theta(0)=0$, $\RDO = 0.1$, $\ZDO = 0$, $\thDO = 0$.  %
}
Relative to a rest observer, the dimensionless Newtonian speed $v(\tau)$ of a test particle on each helicoid is oscillatory but in general tends to unity for large  values of $\tau$. Figure~\ref{fig:wzlayers} displays the evolution of 8 geodesics initially arranged on two different rings, each of radius 0.06. Four of them emanate from a ring lying in  the plane $Z=0.01$ with initial  $\theta$ values of $0,\pi/2, \pi, 3\pi/ $  and four of them from a ring with the same radius and same $\theta$ values lying in the plane $Z=-0.01$. The gravitational wave with helicity 2 is propagating in the negative $Z$-direction. The figure on the right has $\tau$ for all geodesics varying from zero to $2 \times 10^{3}$ while that on the right shows the evolution from zero to  $5\times 10^{3}$. Both figures capture the cycloidal-like modulations and the general uni-directional  emanations induced by geodesics with definite helicity  but different angular locations on identical radius rings above and below the plane $Z=0$. It is also evident that geodesics emanating from the values $\theta=0$ and $\theta=\pi$ initially execute different motions from those starting at $\theta=\pi/2$ and $\theta=3\pi/2$.
   
\WZLAYERS{0.8}{%
	These plots display the evolution of 8 geodesics initially arranged on  two different rings of radius 0.06 (unresolved as the black dots in the figures). Four of them emanate from  a ring lying in  the plane $Z=0.01$  with initial  $\theta$ values of $0,\pi/2, \pi, 3\pi/2$  and 4 of them from a ring  with the same radius  and same $\theta$ values lying  in the plane $Z=-0.01$. The gravitational wave with helicity 2 is propagating in the negative $Z$-direction. The figure on the right has $\tau$ for all geodesics  varying from zero to $2 \times 10^{3}$ while that on the right shows the evolution from zero to  $5\times 10^{2}$. Both figures capture the cycloidal-like modulations and the general uni-directional  emanations induced by geodesics with definite helicity  but different angular locations on identical radius rings above and below the plane $Z=0$.  It is also evident that geodesics emanating from the values $\theta=0$ and $\theta=\pi$  initially execute  different motions from those starting at $\theta=\pi/2$ and $\theta=3\pi/2$. The parameters of the   background gravitational wave metric are $L_{0}=1$, $q_{1}=q_{2} =3/4$, $\Omega=1/20$  and the remaining  initial conditions are $\RDO = 0.1$, $\ZDO = 0$, $\thDO = 0.$ 
	}	
			
Figure~\ref{fig:wringe} displays the evolution of 16 geodesics initially arranged on two different rings, each with radius 1. Eight of them emanate from a ring lying in  the plane $Z=0.245$  with initial  $\theta$ values of  $n\pi/4$ for $n=0,..,7$  and eight of them from a ring  with the same radius  and same $\theta$ values lying  in the plane $Z=- 0.245$. The gravitational wave with helicity 2 is propagating in the negative $Z$-direction.  The proper time $\tau$  for all geodesics varies from zero to $5.8 \times 10^{5}$. With this graphical resolution no cycloidal-like modulations  are visible. It is evident that  uni-directional  emanations  from different initial angular locations  are being induced by harmonic waves  with definite helicity  irrespective of their origins on rings above or below the plane $Z=0$. It also suggests that for a fixed evolution time a relativistic uni-directional bi-jet structure is dominant.									
 
\WRINGE{0.6}{%
	These space-curves displays the evolution of 16 geodesics initially arranged on  two different rings of radius 1. Eight of them emanate from  a ring lying in  the plane $Z=0.245$  with initial  $\theta$ values of  $n \pi/4$ for $n=0,..,7$  and eight of them from a ring  with the same radius  and same $\theta$ values lying  in the plane $Z=- 0.245$. A background gravitational wave with helicity 2 is propagating in the negative $Z$-direction.  The proper time $\tau$  for all geodesics varies from zero to $5.8 \times 10^{5}$. With this graphical resolution no cycloidal-like modulations  are visible.   It is evident that  uni-directional  emanations  from different initial angular locations  are being induced by harmonic waves  with definite helicity  irrespective of their origins on rings above or below the plane $Z=0$. It also suggests that for a fixed evolution time a relativistic uni-directional bi-jet structure is dominant. The parameters of the background gravitational wave metric are $L_{0}=1$, $q_{1}=q_{2}=1/3$, $\Omega=1/20$  and the remaining  initial conditions are  $\RDO = 0.1$, $\ZDO = 0$, $\thDO = 0$. 
	}

The detailed properties of  time-like geodesics in the perturbative domains of these gravitational wave spacetimes depend on the values assigned to a number of their parameters but the emergent general features have been displayed in figures~(\ref{fig:wringa})--(\ref{fig:wringe}). It appears that, in background definite helicity harmonic waves, all time-like geodesics with non-zero initial Newtonian speeds relative to rest-observers evolve into uni-directional modulated helicoids. Being plane-fronted these characteristics are independent of the initial values of $R$ occupied by test matter. Although plane-fronted gravitational waves (like plane-fronted electromagnetic waves) are physical idealisations they offer a possible mechanism for the initiation of uni-directional jet-like patterns when incident on local distributions of matter. In the next section we explore a class of {\it non-stationary}  chiral solutions to the field equations (\ref{DFTF_LAP_EQ}) that are neither plane-fronted nor harmonic gravitational waves. \\

\newpage\quad\newpage
\section{Compact Gravitational Pulses}
\label{sect:compact} 
In \cite{goto_lasers_JPA,goto_lasers_NIMB}, following pioneering work by Synge \cite{synge}, Brittingham \cite{brittingham}, Stewart \cite{stewart} and Ziolkowski \cite{ziolkowski}, we have shown how to construct complex analytic solutions to the Maxwell vacuum equations describing propagating pulse-like electromagnetic fields. These have found direct application for analysing the behaviour of single-cycle laser pulses. One notable feature of such solutions is that, although they do not have a unique direction of propagation (and hence helicity), they are derived from complex solutions with definite chirality.  To construct real spacetime metrics analytically with analogous pulse-like structures, we now outline how the techniques used to construct antisymmetric Maxwell tensors $F$ satisfying\footnote{%
	The linear divergence operator $\delta$  acting on $p$-forms in spacetime is defined by $\star d\,\star $ in terms of the Hodge map $\star$  associated with a metric $g$.%
	} 
$dF=0$, $\delta F=0$ can be generalised to construct symmetric second rank covariant tensors satisfying (\ref{DFTF_LAP_EQ}). Unlike the gravitational wave solutions above, the metric perturbations are compact in all spatial directions and they will be shown to 	describe  non-stationary spacetimes possessing time-like geodesics with  jet-like families of space-curves.\\   
  
If  $\alpha$ is {\it any}  complex (four times differentiable)  scalar field on $\mathcal{U}$ then $\NABLA{\eta} \NABLA{\eta}\,\LAP{\eta}\alpha=\LAP{\eta}\NABLA{\eta} \NABLA{\eta}\alpha$. Hence if  $\LAP{\eta}\alpha=0$, then $\LAP{\eta}\psi_{0}=0$ and $\DIV{\eta}\psi_{0}=0$  are satisfied with\footnote{%
	For any scalar $\alpha$  and metric tensor $g$, $\NABLA{g}\alpha=d\alpha$ is independent of $g$.%
	} 
$\psi_{0}=\NABLA{\eta}d\alpha$. Furthermore since $\NABLA{\eta}$ is the torsion-free Levi-Civita connection, $\psi_{0}$ is symmetric and trace-free with respect to $\eta$. Hence, such complex $\alpha$ in general give rise  to  real   metrics: 
\begin{eqnarray*}
	g_{0} &\;=\;& \eta + \psi_{0}'
\end{eqnarray*}	
with curvature. The particular complex solution $\alpha$ of relevance here is given in the $(T,R,\theta,Z)$ chart $\Phi_{\UU}$ above  as
\begin{eqnarray}\label{PulseSeed}
	\alpha(T,R,Z) &\;=\;& \frac{\kappa}{R^{2}+Q_{12}(T,Z)} 
\end{eqnarray}
where 
\begin{eqnarray*}
	Q_{12}(T,Z) \;\equiv\; \lrb\df Q_{1}+i(Z-T)\rrb\lrb\df Q_{2}-i(Z+T)\rrb
\end{eqnarray*}
with $\kappa,Q_{1},Q_{2}$  strictly positive real dimensionless constants. The scalar $\alpha(T,R,Z)$ is then singularity-free in $T$, $R$ and $Z$ and clearly axially-symmetric with respect to rotations about the $Z$-axis. It also gives rise to an axially-symmetric complex  (zero chirality) tensor $\psi_{0}$ satisfying\footnote{%
	Since $\NABLA{\eta}$ is a flat connection, if $K$ is an $\eta$-Killing vector then the operator  $ \NABLA{\eta}\mathcal{L}_{K} = \mathcal{L}_{K}\NABLA{\eta} $  on all tensors.%
	} 
$\mathcal{L}_{\partial_{\theta}}\psi_{0}=0$. In $\mathcal{U}$, the real axially-symmetric metric tensor $g_{0}$ has  non-zero  components in the coframe  $\CF$:
\begin{eqnarray}
	\nonumber	g_{00} &\;=\;& -1 + \partial^{2}_{TT}\,\alpha^{\prime} \\[0.1cm]
	\nonumber 	g_{01} &\;=\;& g_{10} \;\;=\;\; \partial^{2}_{TR}\,\alpha^{\prime} \\[0.1cm]
	\nonumber	g_{03} &\;=\;& g_{30} \;\;=\;\; \partial^{2}_{TZ}\,\alpha^{\prime} \\[0.1cm]
	\label{PulseMetric} g_{11} &\;=\;& 1+\partial^{2}_{RR}\,\alpha^{\prime} \\[0.1cm]
	\nonumber	g_{13} &\;=\;& g_{31} \;\;=\;\; \partial^{2}_{RZ}\,\alpha^{\prime} \\[0.1cm]
	\nonumber	g_{22} &\;=\;& 1 + \partial_{R}\,\alpha^{\prime}\,/\,R \\[0.1cm]
	\nonumber	g_{33} &\;=\;& 1 + \partial^{2}_{ZZ}\,\alpha^{\prime}
\end{eqnarray}
where $\alpha^{\prime}\equiv\Re\,\alpha$ and satisfies $\mathcal{L}_{\partial_{\theta}}g_{0}=0$. A rest observer field in this spacetime metric is  $\lrb1/{\sqrt{1-\partial^{2}_{TT}\psi_{0}^{\prime} \,}}\rrb\partial_{T}\in T\mathcal{U}$. \\

Complex symmetric tensors $\psi_{m}$ with integer chirality $m>0$ satisfying $\LAP{\eta}\psi_{m}=0$, $\DIV{\eta}\psi_{m}=0$, $\TR{\eta}\psi_{m}=0$ and $\frac{1}{i}\mathcal{L}_{\partial_{\theta}}\psi_{m}=m \psi_{m}$  may be generated from $\psi_{0}$ by repeated covariant differentiation with respect to a particular $\eta$-null  and $\eta$-Killing complex vector field $S$:
\begin{eqnarray}\label{PulsePsi}
	\psi_{m} &\;=\;& \underbrace{\NABLA{\eta}_{S}\cdots\cdots\NABLA{\eta}_{S}}_{\text{$m$ times}}\psi_{0} \qquad\text{where}\qquad \eta(S,-) \;=\; d(\,Re^{i\theta}\,)
\end{eqnarray}
$\text{i.e.}\quad S \;=\; e^{i\theta}\lrb\partial_{R} + (i/R)\partial_{\theta}\rrb$. Solutions with negative integer chirality can be obtained by complex conjugation of the positive chirality complex eigen-solutions. Each $\psi_{m}$ defines a spacetime metric $g_{m}=\eta + \psi'_{m}=\eta + \Re\,\psi_{m}$ on $\mathcal{U}\in\mathcal{M}^{(g_{m})}$ which, for $m\neq 0$, is not axially symmetric:  $\mathcal{L}_{\partial_{\theta}} g_{m}\neq 0$. An indication of the nature of the spacetime geometry determined by $g_{m}$ on $\mathcal{M}^{(g_{m})}$ is given by analysing the structure of the associated Ricci curvature scalar $\CVRR{m}(T,R,Z)$. Unlike in the gravitational wave spacetimes, this scalar is not identically zero. For $m=0$ it is axially symmetric and in the chart $\Phi_{\UU}$ its independence of  $\theta$  means that  for values of a fixed radius its structure can be displayed for a range of  $T$ and $Z$  values given a choice of parameters $(Q_{1},Q_{2},\kappa,L_{0})$. Regions where $\CVRR{0}(T,1,Z)$ change sign are clearly visible in the right side of  figure~\ref{fig:rrzp} where a 2-dimensional density plot shows a pair of prominent loci that separately approach the future ($T\geq0$)  light-cone of the event at $\{R=1,\,T=0,\,Z=0\}$. A more detailed graphical description of $\CVRR{0}(T,1,Z)$ is given in the left hand 3-dimensional plot in figure~\ref{fig:rrzp} where an initial  pulse-like  maximum around $T\simeq 0$ evolves into a pair of enhanced loci with localised peaks at values of $Z$ with {\it opposite signs} when $T\geq 1$. In this presentation the maximum pulse height has been normalised to unity. This characteristic behaviour is similar to that possessed by $\alpha^{\prime}(T,1,Z)$. It suggests that ``tidal forces'' (responsible for the geodesic deviation of neighbouring geodesics \cite{schutz,laemmerzahl,perlick}) are concentrated in spacetime regions where components of the Riemann tensor of $g_{0}$ have pulse-like behaviour in domains similar to those  possessed by $\CVRR{0}(T,R,Z)$. \\

Explicit formulae for $\CVRR{0}(T,R,Z)$ and $\HMAX(T,R,\theta,Z)$ are not particularly illuminating\footnote{%
	Since $g_{0}$ is axially-symmetric, the function $\HMAX$ is independent of $\theta$. 
	}.
However, for fixed values of the parameters $(Q_{1},Q_{2},\kappa,L_{0})$, their values can be plotted numerically in order to gain some insight into their relative magnitudes in any perturbative domain $\mathcal{P}_{\mathcal{U}}$. With $Z$ fixed at zero, figure~\ref{fig:hmaxrrzp} displays such plots as functions of $R$ and a set of $T$ values. It is clear that in perturbative domains the curvature scalar may exceed unity. Since in general:
\begin{eqnarray*}
	\CVRR{0}(T,R,Z) &\;=\;& \mathcal{Q}(T,R,Z)\kappa^{2} + O(\kappa^{3})
\end{eqnarray*}	
where $\mathcal{Q}$ is a non-singular rational function of its arguments and the tensor $h$ is, by definition, of order $\kappa$, figure~\ref{fig:hmaxrrzp} demonstrates that relative tensor $\kappa$-orders are not, in general, indicators of their corresponding relative magnitudes. \\

\HMAXRRZP{0.8}{%
	The axially-symmetric expressions $\vert\CVRR{0}(T,R,0)\vert$ and $\HMAX(T,R,0,0)$ are plotted as functions of $R$ 
	for $T=0,\,0.25,\,0.5,\,0.75$ and parameters $L_{0}=1$, $Q_{1}=Q_{2}=1$, $\kappa=1/4$. Regions where the blue curves lie under the red dotted line denote perturbative regions $\mathcal{P}_{\mathcal{U}}$. The grey shaded regions clearly indicate curvature scalars that are greater in magnitude than unity despite lying within $\mathcal{P}_{\mathcal{U}}$ regions. 
	}

\RRZP{0.9}{%
	An indication of the nature of the spacetime geometry determined by $g_{0}$ on $\mathcal{M}$  is given by the structure of the associated Ricci curvature scalar $\CVRR{0}$. Regions where $\CVRR{0}(T,1,Z)$ change sign are clearly visible in the right side where a 2-dimensional density plot shows a pair of prominent loci that separately approach the future ($T\geq 0$) light-cone of the event at $\{R=1,\,T=0,\,Z=0\}$. A more detailed graphical description of $\CVRR{0}(T,1,Z)$ is given in the  left hand 3-dimensional plot where an initial  pulse-like maximum around $T\simeq 0$ evolves into a pair of enhanced loci with peaks at values of $Z$ with {\it opposite signs} when $T\geq 1$. This Ricci curvature scalar is generated from a metric perturbation pulse with parameters $L_{0}= 1$, $Q_{1}=Q_{2}= 1$, $\kappa = 1/4$.%
	}

The time-like geodesics of $g_{0}$ satisfying (\ref{NORM}) and (\ref{GEOD_EQ}) on $\UUP$ are found numerically following the procedures given in the previous section. To emulate matter on a thick accretion disc test particles are placed  initially on rings with fixed $R$ lying in planes with fixed $Z$ with specified initial conditions in perturbative domains. Since $g_{0}$ is axially symmetric with respect to $Z$, all geodesics will inherit this symmetry. The figures below exhibit time-like geodesics starting at events with $\lrb L_{0}T=0, \,R=\wh{R}(0), \, Z=\wh{Z}(0) \rrb$ and evolving to $\lrb L_{0}T=\taumax, \, R=\wh{R}(\taumax)\right.$,  $\left.Z=\wh{Z}(\taumax) \rrb$ for various values of $\wh{\theta}(0)$. We assign to a collection of evolved time-like geodesics the value of the dimensionless aspect ratio $\AR$ defined by:
\begin{eqnarray*}
	\AR &\;\equiv\;& \left| \frac{\wh{Z}(\taumax)-\wh{Z}(0)}{\wh{R}(\taumax)-\wh{R}(0)} \right|.
\end{eqnarray*}
On the left of figure~\ref{fig:zpzlayers} six geodesics associated with pulse parameters $Q_{1}=Q_{2}=1$ are shown emanating from six locations with $\theta$ values $0, \pi/3, 2\pi/3, \pi, 4\pi/3, 5\pi/3$  on a ring  with radius $10^{-4}$ in the plane $Z=0.735$ and six from similarly arranged points on rings of the same radius at  $Z=0.245$, $Z=-0.245$ and $Z=-0.735$. The initial locations are not resolved in these figures. The 24 geodesics each evolve from $\tau=0$ to $\tau=10^{4}$ and clearly display an axially symmetric  bi-directional jet structure from the rings in conformity  with the expectations based on the spacetime structure of $\CVRR{0}(T,1,Z)$ in figure~\ref{fig:rrzp}. The figure on the right resolves the structure of the directional bi-jet array for $0\leq\tau\leq 100$.  A single uni-directional jet-array arises when only one ring is populated with matter. This figure demonstrates that the jets from the sources at $Z=\pm 0.245$ have a dimensionless aspect ratio $\mathcal{A}(10^{4})=64.7$ much greater than those produced from the sources at $Z=\pm 0.735$ where $\mathcal{A}(10^{4})=3.22$. The  equality of the aspect ratios for a pair of jets produced from sources placed symmetrically around $Z=0$ is a result of the choice $Q_{1}=Q_{2}$. %
\ZPZLAYERS{0.8}{%
	On the left  six geodesics are shown emanating from  six locations with $\theta$ values $0, \pi/3, 2\pi/3, \pi, 4\pi/3, 5\pi/3  $  on a  ring  with radius $10^{-4}$ in the plane $Z=0.735$ and six from similarly arranged points on rings of the same radius at  $Z=0.245$, $Z=-0.245$ and $Z=-0.735$. The initial locations are not resolved in these figures. The 24 geodesics each evolve from $\tau=0$ to $\tau=10^{4}$ and clearly display an axially symmetric  bi-directional jet structure from the rings in conformity  with the expectations based on the spacetime structure of $\CVRR{0}(T,1,Z)$ in figure~\ref{fig:rrzp}. The figure on the right resolves the structure of this jet array for $0\leq\tau\leq 100$. All geodesics are generated with the additional initial  conditions $\RDO = 0$, $\ZDO = 0$, $\thDO = 0.4$ and the background perturbation pulse has parameters $L_{0}=1$, $Q_{1}=Q_{2}=1$, $\kappa = 1/6$.  A single uni-directional jet-array arises when only one ring is populated with matter. This figure demonstrates that the jets from the sources at $Z=\pm 0.245$ have a dimensionless aspect ratio $\mathcal{A}(10^{4})=64.7$ much greater than those produced from the sources at $Z=\pm 0.735$ where $\mathcal{A}(10^{4})=3.22$. %
	}
When $Q_{1}\neq Q_{2}$, it is possible to produce, from a pair of sources on rings arranged symmetrically with respect to $Z=0$,  a pair of uni-directional jets with members of each jet having different aspect ratios. This is demonstrated in figure~\ref{fig:zpzlayersq} where the jet sources on the left have larger separations in $Z$ than those in the centre, all other initial conditions and pulse parameters being the same. If all such sources are in evidence, one produces two pairs of oppositely directed jets, as shown in the right-hand figure, but having different aspect ratios: $\mathcal{A}(10^{4})=178.8$ for $\wh{Z}(0)=-0.735$, $\mathcal{A}(10^{4})=57.35$ for $\wh{Z}(0)=-0.245$, $\mathcal{A}(10^{4})=19.98$ for $\wh{Z}(0)=0.245$ and $\mathcal{A}(10^{4})=105.7$ for $\wh{Z}(0)=0.735$. This figure displays an ``asymmetric'' jet structure with {\it dominant} component belonging to the jet in the left-hand figure having an aspect ratio $\mathcal{A}(10^{4})=178.8$. 

\ZPZLAYERSQ{0.8}{%
	On the left, six geodesics are shown emanating from  six locations with $\theta$ values $0, \pi/3, 2\pi/3, \pi, 4\pi/3, 5\pi/3  $  on a  ring  with radius $10^{-4}$ in the plane $Z=0.735$ and six from similarly arranged points on rings of the same radius at $Z=-0.735$. The initial locations are not resolved in these figures. The 12 geodesics each evolve from $\tau=0$ to $\tau=10^{4}$ and clearly display an axially symmetric uni-directional jet structure from the rings. All geodesics are generated with the additional initial  conditions $\RDO = 0$, $\ZDO = 0$, $\thDO = 0.4$ and the background perturbation pulse has parameters $L_{0}=1$, $Q_{1}=1$, $Q_{2}=3$, $\kappa = 1/6$. The figure in the centre shows an oppositely directed jet evolving from similar initial conditions, but with initial $Z=0.245$ and $Z=-0.245$. The figure on the right displays the ``asymmetric'' jet structure obtained by merging both pairs of sources with {\it dominant} component belonging to the jet in the left-hand figure having aspect ratio $\mathcal{A}(10^{4})=178.8$.%
	}
	
\newpage
\section*{Summary and Concluding Remarks} 
In this article, we have exploited a linearisation of the Einstein vacuum equations about a Minkowski spacetime to construct families of spacetime metrics describing non-stationary perturbed geometries. These metrics are constructed from complex, symmetric, covariant tensors $\psi$ with definite chirality satisfying the tensor equations:
\begin{eqnarray*}
	\LAP{\eta}\psi &\;=\;& 0, \qquad \DIV{\eta}\psi \;=\; 0 \qquadand \TR{\eta}\psi \;=\; 0.
\end{eqnarray*}
The family of solutions in section~\ref{sect:GravWaves} describe plane-fronted gravitational waves of definite {\it helicity} while those in section~\ref{sect:compact} are gravitational pulses of definite {\it chirality}. We have explored the nature of the time-like geodesics in perturbative spacetime domains associated with the lowest allowed chirality solutions in each family. \\

Using suitably arranged massive test particles to emulate a thick accretion disc, together with a particular family of fiducial observers, we have displayed a number of characteristic features of these geodesics in each background metric. Within the context of a non-dimensional scheme, solution parameters can be chosen that result in characteristic spatial jet-like patterns in three-dimensions. These have specific dimensionless aspect ratios relative to well-defined directions of a background harmonic gravitational wave or background gravitational pulse and the corresponding orthogonal subspace. For a helicity $\pm 2$ plane-fronted harmonic wave incident at $T=0$ on a bounded region of matter in the vicinity of the spatial plane $Z=0$ in three-dimensions, parameters ($q_{1}=q_{2}$) can be chosen to yield uni-directional bi-jets: i.e. a {\it pair} of time-like geodesic families with Newtonian speeds approaching the speed of light in the Minkowski vacuum, relative to geodesic observers, for large proper times. At such times, each pair lies in the domain $Z>0$ ($Z<0$) if the wave propagates in the direction of increasing (decreasing) $Z$. For the chirality zero gravitational pulse incident at $T=0$ on a similar disposition of matter, one finds that (with $Q_{1}=Q_{2}$) a {\it pair} of oppositely directed jet-like structures arise: i.e. a pair of time-like geodesic families with Newtonian speeds approaching  terminal values less than the speed of light for {\it both} $Z>0$ {\it and} $Z<0$. For a pulse with $Q_{1}\neq Q_{2}$, we have demonstrated the existence of a pair of uni-directional jet-like structures from particular initial conditions. In all these cases, the structures have well-defined aspect ratios that can be calculated numerically. The propagation characteristics for $T>0$ of the pulse responsible for these jet structures in space is discernible from features of the non-zero Ricci scalar curvature associated with the perturbed spacetime domains. \\

We have also stressed that by linearising only the gravitational field equations and analysing the {\it exact} geodesic equations of motion in perturbative spacetime domains, one can capture the full effects of ``tidal accelerations'' on matter produced by the curvature tensor (and its contractions) associated with the metric perturbations. This opens up the possibility of a gravito-ionisation process whereby extended electrically neutral micro-matter can be split into electrically charged components by purely gravitational forces, leading to modifications of matter worldlines by the presence of Lorentz forces. \\

We conclude that both background spacetime families discussed, separately or in superposition with higher chirality solutions, may offer a non-Newtonian gravitational mechanism for the initialisation of a dynamic process leading to astrophysical jet structures, particularly since it is unlikely that such a phenomenon originates from a unique set of initial conditions. \\

Since the arena involving intense gravitational astrophysical phenomena is wide, direct experimental evidence for broad-band gravitational radiation is still in its infancy. However both exact and linearised plane-fronted wave solutions to the Einstein field equations have long been associated with distant localised matter sources. One might speculate that intense gravitational pulses originate from supernova events or active galactic nuclei. A more ambitious theoretical model than that discussed here, involving perturbations about non-Ricci-flat backgrounds \cite{tuckerGEM}, could be invoked to explain such scenarios. It is worth noting however that although the linearised solutions discussed above (\ref{PulseMetric}) have been formulated in regions of spacetime  lying in the future of a particular focal event (on the hypersurface $T=0$), they can also be used to describe a ``collision scenario'' prior to the formation of such an event. The full spacetime then offers a description having a pair of spatially compact domains with $T<0$, with enhanced curvatures, coalescing to generate a focus of gravitational attraction that may initiate the generation of the accretion disc itself needed to seed astrophysical jets. \\

The nature of possible physical sources for the gravitational backgrounds discussed in this article and the influence of electromagnetic interactions needed to set our results into a particular astrophysical context will be discussed more fully elsewhere. \\[0.6cm]

\section*{Acknowledgements} 
RWT is grateful to the University of Bolton for hospitality and to STFC (ST/G008248/1) and EPSRC (EP/J018171/1) for support. Both authors are grateful to V. Perlick for his comments. All numerical calculations have been performed using Maple 2015 on a laptop. \\[1cm]

\appendix
\setcounter{section}{1}
\section*{Appendix}
This appendix outlines the strategy we have adopted in this paper to solve numerically the geodesic equations (\ref{NORM}) and (\ref{GEOD_EQ}) in spacetimes with the metric $g=\eta + h$. The tensor $h$ is constructed to be $h=\Re(\psi)$ where $\psi$ satisfies:
\begin{eqnarray*}
	\LAP{\eta}\psi \;=\; 0, \qquad \DIV{\eta}\psi \;=\; 0 \qquadand \TR{\eta}\psi \;=\; 0. 
\end{eqnarray*}
For gravitational wave spacetimes $\psi$ is given by (\ref{Psi_GW}) for some choice of integers $\sigma$, $m_{1}$ and $m_{2}$. For gravitational pulse spacetimes $\psi$ is given by (\ref{PulsePsi}) for some choice of positive semi-definite integer $m$ where $\psi_{0}=\NABLA{\eta}d\alpha$ and $\alpha$ is a spatially compact solution satisfying $\LAP{\eta}\alpha=0$. In the context of astrophysical jet modelling, we have chosen (\ref{PulseSeed}) with strictly positive, real, dimensionless parameters $\kappa$, $Q_{1}$ and $Q_{2}$. \\

Equation (\ref{NORM}) can then be solved for the timelike component of $\dot{C}$ in an inertial frame and  substituted into (\ref{GEOD_EQ}).  This enables one to reduce (\ref{GEOD_EQ}) to a system of independent first-order ordinary differential equations with respect to a proper-time independent evolution parameter $\tau$ for the remaining space-like components and their $\tau$-derivatives. A feature of the resulting system is the occurrence of Christoffel symbols (associated with $g$) evaluated on $C$ involving partial derivatives (up to fourth order) of a known function: $W$ for gravitational wave spacetimes or $\alpha$ for gravitational pulse spacetimes. The resulting initial value problem then takes the form:
\begin{eqnarray*}
	\frac{d}{d\tau}\,\,\underline{\Lambda}(\tau) &=& \underline{F}\left(\df\,\underline{\Lambda}(\tau),\, \underline{f}(\tau)\,\right), \qquad \underline{\Lambda}(0) \;=\; \underline{\Lambda}_{\;0} 
\end{eqnarray*} 
where $\underline{\Lambda}(\tau)$ and $\underline{f}(\tau)$ are arrays of different length. Such problems containing known functions $\underline{f}(\tau)$ can be tackled using the Livermore Stiff ODE solver \cite{hindmarsh}. This method has been recently implemented for the numerical integration of ODE systems in the Maplesoft algebraic software programme ``Maple'' as the LSODE option. This programme generates a procedure for numerical output that greatly facilitates the subsequent manipulation of graphical data. \\
	
\newpage
\bibliographystyle{unsrt}
\bibliography{GRAV_PULSE}

\end{document}